\documentclass[12pt]{revtex4}
\usepackage{color}
\usepackage{graphicx}
\begin{document}
\title[Short Title]{RC-circuit-like dynamic characteristic of the magnetic domain wall in ferromagnetic nanowires}
\author{\bf Hong-Guang Piao$^{1}$\footnote{Corresponding author electric mail is hgpiao@ctgu.edu.cn},Dong-Hyun Kim$^{1,2}$\footnote{Corresponding author electric mail is donghyun@chungbuk.ac.kr}, Je-Ho Shim$^{2}$, and Li Qing Pan$^{1}$}
\affiliation{$^1$ College of Science, China Three Gorges University, Yichang 443002, P. R. China\\
$^2$ Department of Physics, Chungbuk National University, Cheongju 361-763, South Korea}
\begin{abstract}
We have investigated dynamic behaviors of the magnetic domain wall under perpendicular magnetic field pulses in ferromagnetic nanowires using micromagnetic simulations.
It has been found that the perpendicular magnetic field pulse can trigger the magnetic domain wall motion, where all the field torques are kept to be on the plane of nanowire strip.
The magnetic domain wall speed faster than several hundreds meters per second is predicted without the Walker breakdown for the perpendicular magnetic driving field stronger than $200~\mathrm{mT}$.
Interestingly, the dynamic behavior of the moving magnetic domain wall driven by perpendicular magnetic field pulses is explained by charging- and discharging-like behaviors of an electrical RC-circuit model, where the charging and the discharging of "magnetic charges" on the nanowire planes are considered.
The concept of the RC-model-like dynamic characteristic of the magnetic domain wall might be promising for spintronic functional device applications based on the magnetic domain wall motion.
\\[0.5cm]
\end{abstract}\maketitle\newpage

Precise control of magnetic domain wall (DW) dynamic behaviors in ferromagnetic nanowires has become one of the important issues for realization of spintronic devices based on the DW
motion\cite{Allwood_device,Piao_diode,Yamanouchi_device,Parkin_device,Hayashi_device}.
In ferromagnetic nanowires, the DW motion generally can be driven by applying an external magnetic field\cite{Allwood_device,Piao_diode} or a spin current\cite{Yamanouchi_device,Parkin_device,Hayashi_device}.
For instance, if an external field is applied along the wire axis to drive a transverse DW, a field torque from a cross product between the external field and the transverse DW magnetization is pointing out of the wire plane, generating a demagnetizing field opposite to the field torque.
The demagnetizing field rotates the transverse DW magnetization on the wire plane, leading to the translational motion of the DW along the wire axis\cite{Leeuw}.
For a stronger magnetic field, where the field torque overcomes the demagnetizing field, an inner spin structure of a simple transverse DW deforms with a structural transformation into an antivortex DW structure\cite{Nakatani,Hayashi,Sang-Koog}.
This phenomenon has been well known as the Walker breakdown and the threshold field is called as the Walker
breakdown field\cite{Walker}.\par

By suppressing the Walker breakdown, the DW speed can be made faster keeping the simple transverse DW structure, which implies that the DW device operation speed can be faster with sustaining a stable transverse DW structure.
Numerous studies have been devoted to exploring a suppression mechanism of the Walker breakdown\cite{Jun-Young,Allwood,Glathe,Akunz-Hz,Soo-Man,Piaoieee,byspinwave,Yanming,PiaoAPL}.
It has been reported that the Walker breakdown can be suppressed by using an underlayer with a strong perpendicular magnetic anisotropy\cite{Jun-Young} by applying an assisting magnetic field along the transverse or perpendicular direction \cite{Allwood,Glathe,Akunz-Hz,Soo-Man}, by modulating the nanowire geometry with a periodic wavy shape\cite{Piaoieee,byspinwave}, or by using a cylindrical nanowire\cite{Yanming,PiaoAPL}.
It has been known that the suppression of the Walker breakdown is directly related to the relative strength of the demagnetizing field compared to the external field torque, where both of them are along the perpendicular (normal) direction of flat (cylindrical) nanowires.\par

In this work, we consider the DW motion triggered by the perpendicular field which is acting not as an assisting field additional to the external magnetic field along the wire axis but {\it as an only driving field}, where the field torque rotating each spin is on the flat nanowire plane all the time.
We report our micromagnetic investigation on dynamic behavior of the DW motion  under perpendicular field pulses with systematic variation of the flat nanowire width and thickness, where RC-circuit-like behavior without the Walker breakdown phenomenon is observed.
\par


We have carried out micromagnetic simulations\cite{oommf} based on the Landau-Lifshitz-Gilbert (LLG) equation\cite{Gilbert}.\par

\begin{equation}\label{e01}
\frac{d\mathbf{M}}{dt} = - |\gamma| \mathbf{M} \times \mathbf{H}_{eff} + \frac{\alpha}{ M_s } (\mathbf{M} \times \frac{d\mathbf{M}}{dt}),
\end{equation}
where $ \gamma $ is the gyromagnetic ratio, $\alpha$ is the Gilbert damping coefficient, $\mathbf{M}$ is the magnetization, and $\mathbf{H}_{eff}$ is the effective field from the total energy in the micromagnetic simulation.\par

The straight ferromagnetic nanowire is set to have a length $l = 2000~\mathrm{nm}$.
The nanowire width $w$ is varied from $20$ to $100~\mathrm{nm}$ and the wire thickness $d$ is varied from $2.5$ to $7.5~\mathrm{nm}$.
The detailed geometry and dimension of the wires are schematically illustrated in the Fig. 1(a).
The external driving field pulse strength $H_{ext}$ is varied from $10$ to
$300~\mathrm{mT}$.
The magnetic field pulse width was set to be $1~\mathrm{ns}$. Both the rise and fall times were set to be 0.1 ns.
The driving pulse field is applied along the normal direction (+z) perpendicular to the nanowire plane, as shown in the Fig. 1(b).
In all simulations, the material parameters of Permalloy are used with an
exchange stiffness coefficient of $13~\times 10^{-12}~\mathrm{J/m}$,
 a saturation magnetization $Ms$ of $8.6~\times
10^5~\mathrm{A/m}$, and zero magnetocrystalline anisotropy.
The unit cell dimension is set to be $2.5 \times 2.5
\times 2.5~\mathrm{nm}^3$ and the Gilbert damping constant $\alpha = 0.01$.
A head-to-head transverse DW is initially prepared at the nanowire center before applying the perpendicular magnetic field pulse.\par


As described in Fig. 1, all spins (dark green arrows) are kept to be on the nanowire plane ($xy$-plane), even after the external field torque is exerted, as denoted by the dotted black arrows in the figure.
Based on the LLG equation, it is found that the external field torque is $\vec{\tau}_H = \mathbf{M} \times \mathbf{H_{eff}}$, so that the magnetization direction is determined by
$\frac{d\mathbf{M}}{dt} = - |\gamma| \mathbf{M} \times \mathbf{H}_{ext}$. The demagnetizing field torque is
$\vec{\tau}_d = \mathbf{M} \times \mathbf{H_{demag}}$, so that
$\frac{d\mathbf{M}}{dt} = - |\gamma| \mathbf{M} \times \mathbf{H}_{demag}$ (see Fig. 1(c)).
Under a perpendicular magnetic field, inner spins of the DW are rotated by the external field torque along counterclockwise direction, effectively moving the transverse DW to the left ($-x$) direction along the nanowire, as shown in Fig. 1(b).
Due to the counterclockwise rotating torque on all spins lying on the wire axis, a gradual change of magnetization generates magnetic source charges acting like positive 'N' and negative 'S' "magnetic charges"\cite{Hubert} at the wire edges, as denoted by magenta (N) and cyan (S) colors at the wire edges in the Fig. 1(b).
By the generation of magnetic charges, demagnetizing fields are generated along the $-y$ ($+y$) direction on the left (right) side of the DW, as denoted in the Fig. 1(c).
In the meantime, another torque (dotted black arrows) is generated along the $-z$ direction by the demagnetizing field from the magnetic charges on the wire edges, leading to the accumulation of magnetic charges on the upper and lower flat planes of the nanowire by the magnetization tilting toward the $z$ axis direction, as denoted by 'N' (magenta) and 'S' (cyan) poles in Fig. 1(c). Due to the magnetic charge accumulation, another demagnetizing field is generated toward the $-z$ direction as depicted in the Fig. 1(d).
After the magnetic field pulse is applied, the role of the driving field is replaced by the demagnetizing field from the accumulation of magnetic charges on the wire planes, resulting in the DW motion toward the right ($+x$) direction in the nanowire, as shown in Fig. 1(d).\par

To study the DW dynamic behavior driven by the perpendicular field, we have analyzed magnetization with respect to the time, as illustrated in the Fig. 2(a), where time-dependent magnetization profiles are plotted together with a temporal profile of the perpendicular magnetic field pulse.
In the figure, magnetizations are plotted for a field with 50-mT strength, 1-ns duration time, and 0.1-ns rise/fall time. The nanowire width, w = 50 nm and the thickness, d = 5 nm.
It is observed that the $y$-component magnetization $M_{y}$ does not significantly change under the pulse field driving since $M_{y}$ mainly comes from the magnetization of the structure-stable transverse DW, which is also evidenced by the negligible change of exchange energy curve observed in Fig. 2(b). The total energy is mainly dominated by the energy gain in the Zeeman energy when the pulse field is on.
While the perpendicular pulse field is on, the transverse DW is simply translationally displaced along the wire with keeping the inner spin structure stable, as confirmed in the Fig. 2(c).
Conversely, the time-dependent $M_{z}$ profile is significantly changed under the pulse field driving, temporally consistent with the applied field pulse, which means that the demagnetizing field is almost instantaneously generated toward $-z$ direction and kept to be nearly constant for the field pulse duration, as also observed in the demagnetization energy curve in Fig. 2(b).
It should be noted that the demagnetizing field generated toward $-z$ direction is originated from the magnetic charges accumulation on the nanowire plane, as described in Fig. 1(d).\par

$M_{x}$ profile exhibits a distinctive behavior compared to other magnetization components.
Moreover, very interestingly, after the pulse field is off, the DW gradually moves back to the original position again with keeping the stable spin structure.
As vividly demonstrated in the Fig. 2(a), careful analysis reveals that the time-dependent DW position profile (open square), determined by processing the simulated DW images (see Fig. 2(c)), is exactly matching with the time-dependent $M_{x}$ profile (red solid line).
In case of Fig. 2, the maximum displacement of the DW under the perpendicular field pulse is 146 nmwhere the stable inner spin structure is kept. .
In the case of Fig. 2, the average speed of the DW motion is about 134 m/s and could be faster than when applied field pulse becomes stronger.
Under the pulse field with 200-mT strength and 1-ns duration time, the average DW speed is observed to be about 433 m/s, where the inner structure of the transverse DW is still stably maintained.
Reminding that the time-dependent DW position curve has the same profile shape with the time-dependent $M_{x}$ curve, the change of the instantaneous slope from the steep one to the more moderate one, looks like the curve of an exponential function. If so, the DW speed is expected to be gradually decelerated during the pulsed time.\par

To examine the relation between the time-dependent DW position profile and the driving field pulse, we applied a sequential multiple field pulses perpendicular to the wire plane.
Where 5-times field pulses with 1-ns duration time and 100-mT strength are successively applied.
It is found that the profile of time-dependent $M_{x}$ and DW position present a kind of continuous behavior, as shown in the Fig. 3(a).
These disconnected profiles seem to be quite coherent and the slope change is continued even after new field pulses are applied.
One can think that the DW motion driven by the next field pulse depends on the DW motion driven by the previous field pulse, suggesting that the continuous accumulation of certain physical quantities -- "magnetic charge".\par

Based on above confirmations and the exponential-curve-like profile shape, we are allowed to assume that the phenomenon is related to a charging/discharging effect of the magnetic charge.
The assumption is thought to be valid since there exist magnetic charges on the wire planes and the magnetic charges might behave in a similar way to the electrical charges.
In order to prove our assumption, we consider the Thiele's equation based on the LLG equation\cite{Shim_Thiele}:
\begin{equation}\label{e02}
-\mathbf{\hat{G}} \times \mathbf{\dot{X}}-\mathbf{\hat{D}} \cdot \mathbf{\dot{X}} + k \mathbf{X} = u (\mathbf{\hat{y}} \times \mathbf{H_{ext}} ),
\end{equation}
where, $\mathbf{\hat{G}}$ is a gyrovector and $\mathbf{X}$ is a DW position. $\mathbf{\hat{D}}$ is a damping tensor (negative value) related to the sample geometry, the DW width, and the Gilbert damping constant\cite{Shim_New}. $k$ and $u$ are a stiffness coefficient and a constant, respectively, depending on the sample geometry (demagnetizing factor)\cite{Guslienko2001,Guslienko2006}. $\mathbf{H_{ext}}$ is an external magnetic field define to be along the $z$-direction.
We have found that the DW dynamic behavior is very similar to the one-dimensional (1D) version of dynamic magnetic vortex core behavior in the ferromagnetic nanodisk without the gyroforce term. Therefore, we try to use the 1D Thiele's equation $-\mathbf{\hat{D}} \cdot \mathbf{\dot{X}} + k \mathbf{X} = u (\mathbf{\hat{y}} \times \mathbf{H_{ext}} )$ to explain the observed behavior in the present study.
According to the 1D Thiele's equation, the time-dependent DW position function $x[t]$ follows a relation as $x[t] = C_1 \frac{uH_{ext}}{k}(1-e^{-k t/|D|})$, where $C_1$ is a constant.
It should be noted that the resultant function form is very similar to the charging function form $A_{0}[1 - exp(-t/\tau)]$ in the RC-circuit model\cite{RC-circuit}.
The role of $|D|/k$ seems to be corresponding to an RC characteristic time $\tau$ and the $\frac{uH_{ext}}{k}$ corresponds to $A_{0}$.
\par

In addition, we have also found that the exponential behavior of the DW position does not significantly depend on the rise/fall time, as shown in the Fig. 3(b), where two $M_{x}$ profiles with different rise time (0 and 0.2 ns) are compared.
Both profiles are almost same and almost exactly matching if shifted by the difference of rise time. The change of field strengths does not distort the exponential behavior, either.  Thus, it is concluded that the exponential behavior of the DW position is not originated from the field parameters such as rise/fall time and field strength.
It is also found that the exponential behavior of the DW position substantially depends on the Gilbert damping constant, as demonstrated in the Fig. 3(c), since the RC-characteristic-time-like $|D|/k$ term is a function depending on the Gilbert damping constant\cite{Shim_New}.
Every aspect of the observed exponential behavior seems to let us firmly believe that the RC-like DW dynamic behavior is well explained by charging and discharging of magnetic charges on the nanowire.\par

According to the RC-circuit model, the nanowire itself can be considered as a "magnetic capacitance", where the capacitance $C_m$ is proportional to the area of the wire planes and inversely-proportional to the wire thickness.
To check our scenario, firstly, we have analyzed the fully charged state corresponding to the state with the maximum DW displacement. We simply assume that the present ferromagnetic nanowire system is analogous to the electrical RC-circuit system, where an RC characteristic time constant $\tau$ and the $M_{x}/Ms$ profile are fitted with an RC-circuit-like fitting function of $A_{0}[1 - exp(-t/\tau)]$ for charging and $A_{0} exp (-t/\tau)$ for discharging. $A_{0}$ represents the saturated state and thus, corresponding to the state with the maximum DW displacement.
The fitting is found to be extremely well matching for both the DW position and $M_x/Ms$, as demonstrated in Fig. 3(d), where three fitting curves in the cases of different field pulse strengths of 20, 50, and 100 mT. Note that the profiles for the case of static (not pulsed) applied field become continuously exponential in the figure.
From the fitting, $A_{0}$ values are determined to be 0.06, 0.16, and 0.33 for cases of 20, 50, and 100-mT perpendicular field respectively, which indicates that the magnetic charge at the fully charged state ($Q_m$) is proportional to the applied field strength ($H$) with a coefficient of $C_m=\frac{Q_m}{H} \approx 3\ \mu_B\mathrm{{\AA}}^{-1}\mathrm{T}^{-1}$\cite{Magnetic_charge_unit}.
Moreover, we have found that the characteristic times for different field strengths are $\tau \approx$ 0.79 (20 mT), 0.80 (50 mT), and 0.83 ns (100 mT) become almost same regardless of the field strength.
Considering $\tau$ mainly depends on the wire geometry, dimension, and material parameters in the magnetic RC-circuit system, the negligible change of $\tau$ with different field strengths is well explained.\par

In Fig. 4(a)-(c), the time-dependent DW position profiles are plotted with various widths and thicknesses of nanowires under the magnetic field pulse of 50-mT strength and 0.1-ns rise time. The widths are varied to be 20, 50, and 100 nm and the thickness is changed to be 2.5, 5.0, and 7.5 nm.
In case of the same wire width, the DW maximum displacement becomes smaller for thicker wires. Quantitative analysis reveals that the wire thickness is inverse-proportional to the magnetic capacitance, which is explainable by the fact that the thicker wire has the lager separation for a plate capacitor.
In a similar way, in case of the same wire thickness, the DW maximum displacement becomes lager for wider wires, where the wire width is confirmed to be proportional to the magnetic capacitance. The observed behavior is again explainable by the fact that the wider wire has the wider plane area for a plate capacitor.
Therefore, the relation between the magnetic capacitance and the wire dimension could be simply deduced as $C_m \propto (w\times l)/d$.
If $d$, $l$, and $w$ are fixed, the capacitance will be a constant, as $C_m \approx 3\ \mu_B\mathrm{{\AA}}^{-1}\mathrm{T}^{-1}$ for the case of the Fig. 3(d).\par

Considering the relationship between the capacitance $C_m$ and the wire dimension ($d$ and $w$), we have further analyzed the maximum displacement of the DW motion under static perpendicular fields with variation of the field strength from 10 to 200 mT.
Reminding that the displacement of the DW position is directly related to the generation of magnetic charges on the wire plane,the magnetic static field is considered to act like an energy potential as in the RC-circuit model.
Interestingly, we have observed that the slope of the maximum displacement profile with respect to the static field is constant regardless of field strengths for nanowires with various dimensions, as plotted in the Fig. 4(d). This also matches with the RC characteristic time is mainly determined by R and C, not by the electrical potential. In magnetic analogy, the characteristic time $\tau$ is also constant for each wire regardless of the field strengths (acting like the potential).
To further confirm the RC-circuit characteristic of the DW motion under perpendicular pulse fields, the resistance $R$ is calculated by the relation function $\tau = RC$ for each wire dimension.
We have found that the $R$ value also becomes nearly constant for each wire, not depending on the external field strength, as shown in Fig. 4(e), whereas, under driving field stronger than 100 mT, the $R$ value gradually increases, in particular, fora relatively narrower and thinner wire ($w$ = 20 nm and $d$ = 2.5 nm). This deviation might be explained considering that a stronger perpendicular field modifies the DW width, thereby changing the RC-characteristic-time-like $|D|/k$ term which depends on the DW width. However, our simple RC-circuit-like model is found to be valid within a practical range of the field strengths not stronger than 100 mT.
change.\\par


In conclusion, RC-circuit-like DW dynamics is discovered in ferromagnetic flat nanowires under a perpendicular field pulse.
With a systematic variation of the wire width and thickness as well as the field pulse strength, the RC-circuit-like dynamic DW behavior is stably observed without occurence of the Walker breakdown.
"Magnetic charges" accumulated on the wire plane seem to well explain the observed RC-circuit-like DW dynamic behavior, where the displaced DW back and forth are exactly corresponding to the charging and discharging in the RC-circuit.
The observed RC-circuit-like behavior is well explained as well based on a simple 1D Thiele's equation based on the LLG equation.
Our study opens a new possibility to control the fast DW motion avoiding the Walker breakdown in ferromagnetic nanowires, where the DW dynamics is exactly following the well known RC-circuit-like behavior, which is expected to be promising in designing spintronic logic devices based on the DW motion along ferromagnetic nanowires.\par

\textbf{Acknowledgment} This work was supported by the National Science Foundation of China (Grant Nos. 11474183 and 51371105) and the Korea Research Foundation (NRF) (Grant Nos. 2010-0021735). This research was also supported in part by the Leading Foreign Research Institute Recruitment Program [Grant No 2010-00471]. \par

\textbf{Author Contributions} H.G.P. designed research blueprints, did the simulation and wrote the manuscript. D.H.K. contributed to project design, theoretical analysis, manuscript writing and whole project supervision. J.H.S. and L.Q.P. contributed to the phenomenon analysis. All authors have commented on the manuscript.\par

\textbf{Author Information} The authors declare no competing financial interests.
Correspondence and requests for materials should be addressed to D.H.K (donghyun@chungbuk.ac.kr) or H.G.P. (hgpiao@ctgu.edu.cn).
\newpage
\newpage

Figure 1. (Color online) (a) Schematics of a head-to-head transverse DW spin structure and a ferromagnetic flat nanowire dimension under zero field.
(b) Schematics of the magnetic charge (N and S) accumulation at the wire edges by the external field torque (dotted black arrows) denoted as the magenta (N) and cyan (S) colors under the perpendicular pulse field ($H_{ext}$). (c) Schematics of the magnetic charge (N and S) accumulation at the wire planes by the torque of the demagnetizing field (hollow blue arrows) along the transverse direction of the wire.(d) Schematics of the demagnetizing field generation along the -z-direction after the field pulse is off. The rotating directions (magnetization direction) of all spins are denoted as green arrows.
\\[1cm]

Figure 2. (Color online) (a) Time-dependent magnetization and DW position profiles. $H_{ext} = 50~\mathrm{mT}$ field pulse profile is on the bottom. (b) Profiles of Zeeman, demagnetization, exchange, and total magnetic energies. (c) DW positions at $t = 1$, $2$, and $5~\mathrm{ns}$ in the nanowire of $w = 50~\mathrm{nm}$, $d = 5~\mathrm{nm}$. Color code on the right represents the $M_{y}$ component.
\\[1cm]

Figure 3. (Color online) (a) Time-dependent $M_{x}/Ms$ and DW position profiles (up) under multiple sequential perpendicular field pulses with $H_{ext} = 100~\mathrm{mT}$ (bottom). (b) $M_{x}/Ms$ profiles with rise times of 0 and 0.2 ns. The field pulse profiles are denoted as dotted line. (c) $M_{x}/Ms$ profiles with different damping parameters of 0.01, 0.05, and 0.1. (d) DW positions under pulse fields (solid) and static fields (open) with 20, 50, and 100-mT field strengths. Fitting curves (solid line) for time-dependent $M_{x}/Ms$ are plotted together.
\\[1cm]

Figure 4. (Color online) DW positions with variation of wire width and thickness. The thickness is varied to be 2.5, 5.0, 7.5 nm for (a) the width of 20 nm (inset is a zoomed-in figure corresponding to the dotted box region), (b) 50 nm, and (c) 100 nm. (d) The maximum DW displacement with respect to the static field strength and (e) the resistance $R$ profile with variation of width and thickness, fixing the field pulse strength in each case. The fitting curves are denoted as solid lines.
\newpage

\begin{figure}[!h]\label{f001}
\setlength{\abovecaptionskip}{6pt}\centering
\includegraphics[width=0.8\textwidth]{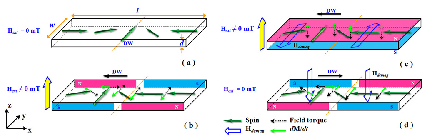}\\FIG. 1
\end{figure}\newpage

\begin{figure}[!h]\label{f002}
\setlength{\abovecaptionskip}{10pt}\centering
\includegraphics[width=1\textwidth]{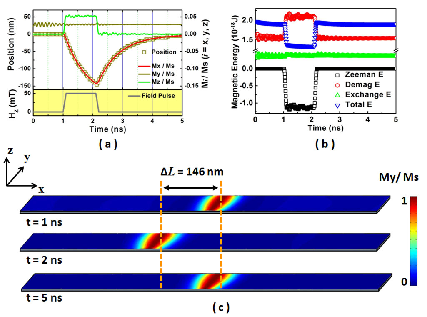}\\FIG. 2
\end{figure}\newpage

\begin{figure}[!h]\label{f003}
\setlength{\abovecaptionskip}{10pt}\centering
\includegraphics[width=0.8\textwidth]{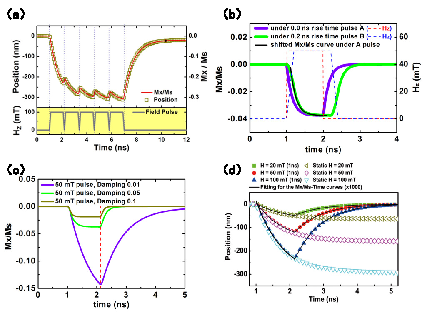}\\FIG. 3
\end{figure}\newpage

\begin{figure}[!h]\label{f004}
\setlength{\abovecaptionskip}{10pt}\centering
\includegraphics[width=1\textwidth]{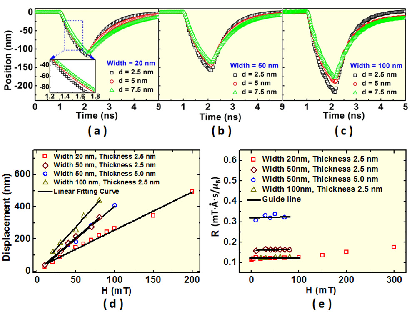}\\FIG. 4
\end{figure}\newpage
\end{document}